# Low-Temperature Eutectic Synthesis of PtTe$_2$ with Weak Antilocalization and Controlled Layer Thinning

*Song Hao, Junwen Zeng, Tao Xu, Xin Cong, Chenyu Wang, Chenchen Wu, Yaojia Wang, Xiaowei Liu, Tianjun Cao, Guangxu Su, Lanxin Jia, Zhangting Wu, Qian Lin, Lili Zhang, Shengnan Yan, Mengfan Guo, Zhenlin Wang, Pingheng Tan, Litao Sun, Zhenhua Ni, Shi-Jun Liang[*], Xinyi Cui and Feng Miao[*]*

Dr. S. Hao, J. W. Zeng, C. Y. Wang, C. C. Wu, Y. J. Wang, X. W. Liu, T. J. Cao, G. X. Su, L. X. Jia, L. L. Zhang, S. N. Yan, Prof. Z. L. Wang, Dr. S. J. Liang, Prof. F. Miao
[1]National Laboratory of Solid State Microstructures, School of Physics, Collaborative Innovation Center of Advanced Microstructures, Nanjing University, Nanjing 210093, China
Corresponding authors: miao@nju.edu.cn; sjliang@nju.edu.cn
Dr. T. Xu, Prof. L. T. Sun
[2]SEU-FEI Nano-Pico Center, Key Laboratory of MEMS of Ministry of Education, Southeast University, Nanjing 210096, China
X. Cong, Prof. P. H. Tan
[3]State Key Laboratory of Superlattices and Microstructures, Institute of Semiconductors, Chinese Academy of Sciences, Beijing 100083, China
Dr. Z. T. Wu, Q. Lin, Prof. Z. H. Ni
[4]School of Physics, Southeast University, Nanjing 211189, China.
M. F. Guo, Prof. X. Y. Cui
[5]State Key Laboratory of Pollution Control and Resource Reuse, School of the Environment, Nanjing University, Nanjing 210046, China

Keywords: Eutectic Synthesis; Type-II Dirac semimetal; Platinum Tellurium; Weak Antilocalization; Layer-by-layer thinning

Metallic transition metal dichalcogenides (TMDs) have exhibited various exotic physical properties and hold the promise of novel optoelectronic and topological devices applications. However, the synthesis of metallic TMDs is based on gas-phase methods and requires high temperature condition. As an alternative to the gas-phase synthetic approach, lower temperature eutectic liquid-phase synthesis presents a very promising approach with the potential for larger-scale and controllable growth of high-quality thin metallic TMDs single crystals. Herein, we report the first realization of low-temperature eutectic liquid-phase synthesis of type-II Dirac semimetal PtTe$_2$ single crystals with thickness ranging from 2 to 200 nm. The electrical measurement of synthesized PtTe$_2$ reveals a record-high conductivity of as high as 3.3×10$^6$ S/m at room temperature. Besides, we experimentally identify the weak antilocalization behavior in the type-II Dirac semimetal PtTe$_2$ for the first time. Furthermore, we develop a simple and general strategy to obtain atomically-thin PtTe$_2$ crystal by thinning as-synthesized bulk samples, which can still retain highly crystalline and exhibits excellent electric conductivity. Our results of controllable and scalable low-temperature eutectic liquid-phase synthesis and layer-by-layer thinning of high-quality thin PtTe$_2$ single crystals offer a simple and general approach for obtaining different thickness metallic TMDs with high-melting point transition metal.

## 1. Introduction

Two-dimensional (2D) semiconducting transition metal dichalcogenides (TMDs) materials have been widely studied for promising applications in electronic [1, 2], optoelectronic [3, 4], spintronic devices [5, 6], plasmonics [7, 8] and membrane technology [9, 10]. Beyond semiconducting TMDs, metallic TMDs materials with exotic properties have also attracted much attention over the past few years. [11-17] However, synthesis and properties of metallic TMDs materials remain less unexplored compared with well-studied semiconducting TMDs, although they show enormous potential applications, such as improving contact properties of 2D semiconductor transistors [18, 19] and enhancing catalysis performances. [20-23]

Recently, platinum tellurium (PtTe$_2$) has been identified as a type-II Dirac semimetal through angle-resolved photoemission spectroscopy (ARPES), [24] which opens a new door to investigate various intriguing physical properties of metallic TMDs.

In stark contrast to $Cd_3As_2$ (known as type-I Dirac semimetal), $PtTe_2$ has a tilted Dirac cone with linear dispersion. [25-28] The titled Dirac cone at certain momentum direction may provide a new platform to investigate anisotropic magneto-transport properties, similar to type-II Weyl semimetals. [29-32] According to ARPES and theoretical results, [24] the Dirac cone is buried in the valence band and Te-p orbitals are responsible for formation of valence bands. The overlap repulsion between filled 5p orbitals of Te is much stronger in each Te-atom sheet than adjacent two Te-atom sheets, creating an ultra-strong covalent bonds joining adjacent layers. [33] The strong interlayer interactions in $PtTe_2$ would lead to many interesting thickness-dependent electronic properties. [34-37] The $PtTe_2$ with different thicknesses can be accessed either by direct chemical vapor deposition or layer-by-layer thinning approach. In particular for the controlled layer thinning approach, it is very promising to be a simple, general and nondestructive strategy to achieve metallic TMDs with strong interlayer interactions. Besides, the heavy Pt and Te atoms in type-II Dirac semimetal $PtTe_2$ lead to a strong spin-orbital interaction, which enables the occurence of weak antilocalization. To investigate these peculiar physical properties, it is highly desirable to synthesize high-quality stoichiometric $PtTe_2$ single crystals with different thicknesses through a simple and general approach.

Up to now, chemical vapor transport (CVT) technique and chemical vapor deposition (CVD) are mostly used methods to synthesize single crystals. However, CVT-synthesized samples often encounter nonstoichiometric issues. [35, 38] Moreover, the strong interlayer interactions would lead to a challeng in obtaining atomically-thin crystals through mechanical exfoliation of CVT-synthesized samples, [33-35] or through direct chemical vapor deposition, which drastically affects the studies of fundamental properties and potential topological device applications of PtTe$_2$ thin crystals. Even worse, both CVT or CVD techniqe requires high temperature sublimation of precurses to synthesize the metallic TMDs with high melting-point transition metal. From the viewpoint of thermodynamics, it is noteworthy that elements Pt and Te can form eutectic solution and yield PtTe$_x$ compounds with various stoichiometric ratio, dependent on reaction temperature and atomic ratio of precursors. [39] Therefore, it is quite promising to develop a low-temperature synthesis of stoichiometric and high-quality thin PtTe$_2$ single crystals through the facile eutectic solidification, which can be used to uncover the weak antilocalization phenomenon in type-II Dirac semimetals.

In this work, we report the first low-temperature eutectic liquid-phase synthesis of large scale PtTe$_2$ single crystals with thickness ranging from 2 to 200 nm, by utilizing a novel solid-liquid-solid approach. Electrical measurement of PtTe$_2$ four-terminal Hall-bar devices reveals an ultra-high electronic conductivity of $3.3 \times 10^6$ S/m at room temperature, which is a record high among metallic TMDs reported so far. We also experimentally identify the weak antilocalization in type-II Dirac semimetals. Furthermore, we develop a simple and general layer-by-layer thinning method to obtain atomically-thin PtTe$_2$ crystal, which can retain the excellent electric conductivity similar to as-grown samples.

**2. Results and Discussion**

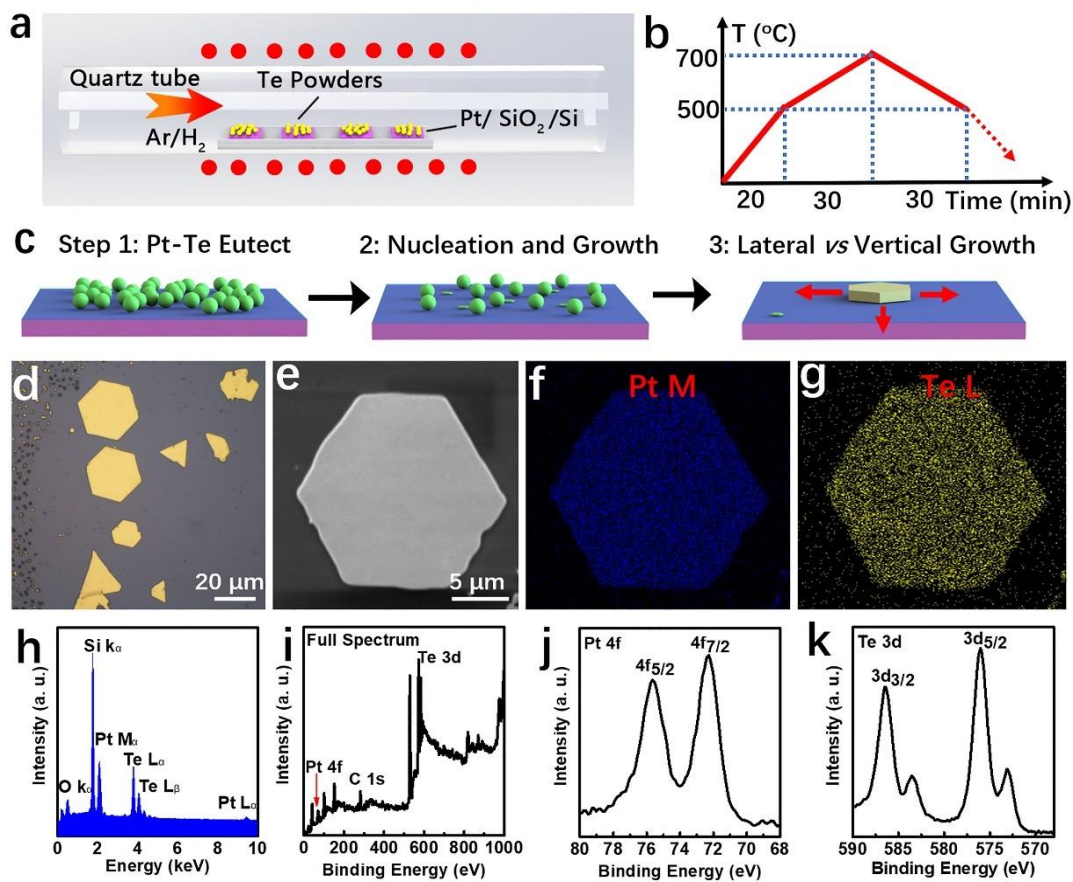

**Figure 1.** Direct synthesis of PtTe$_2$ single crystals on SiO$_2$/Si substrate in an atmospheric-pressure CVD-like furnace. (a) A schematic diagram of CVD-like system. The corresponding locations of Te powders and Pt/SiO$_2$/Si substrates are marked. (b) Temperature programming processes of Te powders. (c) Schematic flow chart of the formation of PtTe$_2$ single crystals. Pt films can be slightly dissolved into Te when exceeding melting point of Te, then Pt-Te eutectic undergoes a nucleation and growth processes, yielding PtTe$_2$ single crystals. (d) The optical microscopy image of typical hexagon and triangle shape PtTe$_2$ single crystals. (e-h) The SEM and corresponding EDS mapping images of Pt and Te and elemental analysis of the hexagonal PtTe$_2$ flake. (i-k) The corresponding XPS spectra of representative sample.

We synthesized PtTe$_2$ on amorphous SiO$_2$/Si substrates in a CVD-like furnace through a novel and facile eutectic solidification method, as shown in Figure 1a. Due to lower vapor pressure of PtO$_2$ and lower reaction activity of PtCl$_2$, we selected elemental Pt as the reactant precursor and SiO$_2$/Si as substrate, which is different from previous works of using powdered reactant as precursors to synthesize metallic TMDs on mica substrate. [14, 19] Based on phase diagram of Pt-Te system (see Figure S1 in the Supporting Information), we design a programing temperature of furnace as shown in Fig. 1b, [39] and demonstrate the whole growth processes in Figure 1c. The principle of chemical reaction is based on the Pt-Te eutectic solidification. According to the phase diagram (see Figure S1), the excess Te powders can react with Pt to produce the pure phase $PtTe_2$. The reaction process can be described in the following way. When temperature is close to 400 °C, as-deposited Pt films covered with Te powders would gradually dissolve into excess liquid Te, which results in formation of Pt-Te solid solution. As the reaction temperature continue to increase, excess liquid Te would be evaporated and carried by carrier gas to downstream, providing a Te-rich atmosphere to prevent as-grown crystals from oxidation. Eventually, the crystals are formed after nucleation and growth, followed by lateral and vertical competitive expansion of crystalline domains. By analogy with vapor-liquid-solid (VLS) growth mechanism for semiconducting nanowires, the solid-liquid-solid (SLS) growth is proposed to account for the growth mechanism, in which the whole reaction proceeds continuously from Pt-Te solid, liquid to solid. [40] In the synthesis of $PtTe_2$ single crystals, Te not only serves as reactant precursor, but also lowers the melting point of elemental Pt through forming eutectic system similar to prior works where tellurium also severs as a flux and is removed by the $Ar/H_2$ flow finally. [41, 42] Note that the term CVD is not strictly suitable for the chemical reaction described above, because precursors react in a molten solution rather than in the vapor phase. [43-45]

Based on the SLS growth mechanism, as-synthesized samples display either hexagonal or triangle shape (see Figure 1d), with lateral size of several tens of micrometers. The uniform color contrast both for hexagonal and triangular PtTe$_2$ single crystals indicates relatively high quality of as-grown samples. It is noteworthy that a sea of small particles observed in the left area of optical microscopy image can be ascribed to Pt-Te byproducts under facile growth conditions, The good separation of small particles with PtTe$_2$ single crystals ensures the investigation of physical property of pure-phase PtTe$_2$ single crystals (see Figure S2 in the Supporting Information). To identify chemical elements and analyze atomic ratio of as-grown samples, we carried out elemental mapping using energy-dispersive X-ray spectroscopy (EDS) over the whole region of PtTe$_2$ single crystal (see Figure 1e). EDS mappings of Pt and Te (see Figure 1f and g) confirm the uniform distribution of Pt and Te elements in PtTe$_2$, which is consistent with optical microscopy result. The atomic ratio of Pt to Te determined by EDS spectrum (see Figure 1k) is ~1:1.95, fairly close to its nominal stoichiometry. The XPS survey spectrum (see Figure 1i) shows the predominant signals of Pt and Te elements. Pt 4f and Te 3d narrow scans in XPS spectra (see Figure 1j and k) exhibit characteristic sub-peaks centered at ~75.6 and ~72.3 eV below Fermi level, assigned to doublet Pt $4f_{5/2}$ and $4f_{7/2}$ of PtTe$_2$. Based on optical microscopy result, the peaks located at ~583.6 and ~573.1 eV are assigned to Te $3d_{3/2}$ and $3d_{5/2}$ of PtTe$_2$, in accordance with previous results. [23]To investigate whether the as-grown samples are pure phase PtTe$_2$ or mixed phase of PtTe$_2$ and Pt$_2$Te$_3$, we carried out the X-ray diffraction (XRD) experiment on one of as-grown Hexagonal samples, with results shown in Figure S3 (see Supporting Information). Peaks appearing in 7.84º, 19.57º, 33.66º and 39.35º correspond to (001), (102), (203) and (213) of PtTe$_2$ (JCPDS card No.18-0977), respectively. The other two peaks at 13.00° and 37.14° can be well indexed to (111) and (422) silicon from substrate (JCPDS card No.27-1402), respectively. The results clearly show the high purity of hexagonal PtTe$_2$ single crystals. To further confirm the PtTe$_2$ pure phase of as-grown samples, we performed additional XRD measurements of as-grown samples on other Si substrates (label as sample 1, sample 2 and sample 3) and different areas on same Si substrates (label as sample -1, sample -2 and sample -3), with results shown in Figure S4 (see Supporting Information). A number of peaks (marked by five-pointed star) in the XRD patterns can be well-indexed to PtTe$_2$ (JCPDS card No.18-0977) and the other two peaks marked by asterisk are assigned to silicon. Note that the peaks associated with Pt$_2$Te$_3$ are not observed in the XRD patterns. Considering all these evidences above, we conclude that the excess Te powders can react with Pt to produce pure phase PtTe$_2$, instead of co-existence of PtTe$_2$ and Pt$_2$Te$_3$.

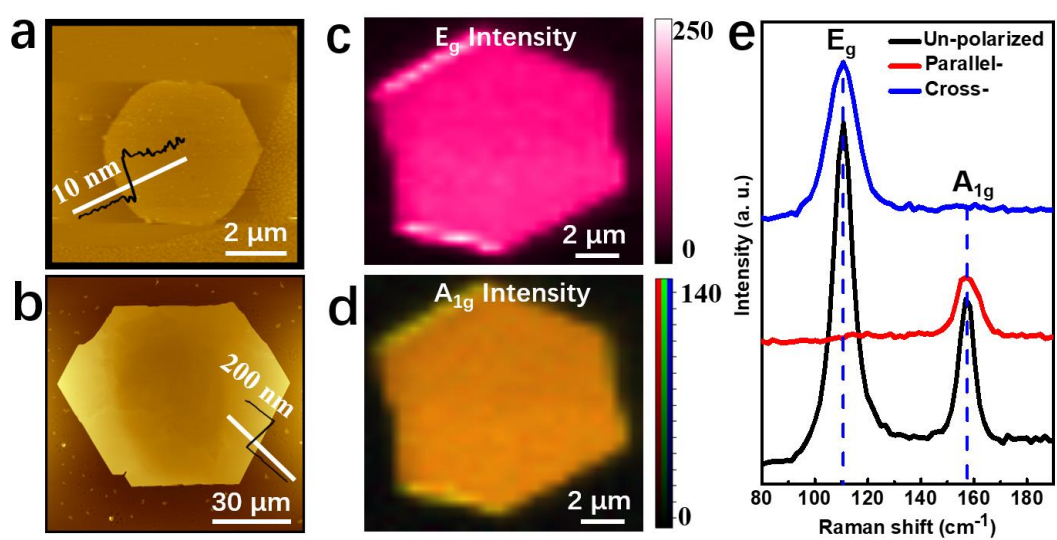

**Figure 2.** (a, b) The AFM topography images of typical samples with different thicknesses. The inserted height profiles in (a, b) are taken along the white solid lines. (c, d) Raman intensity mappings of E$_g$ and A$_{1g}$ characteristic modes for PtTe$_2$ sample. (e) Raman spectra measured at angle of 0° under un-, parallel, and cross-polarized configurations.

To obtain information about thickness, lateral size and structure of as-grown samples, we performed measurements of AFM and Raman spectroscopy/Mapping. Figure 2a and b exhibit two hexagonal $PtTe_2$ single crystals of different lateral sizes, corresponding to thicknesses of ~10 nm and ~200 nm, respectively. It should be emphasized that atomically-thin $PtTe_2$ single crystals with lateral size of several hundred nanometers can also be achieved, which is comparable to that of CVD-grown $PtSe_2$ single crystal, [35] by reducing the thickness of pre-deposited Pt films and shortening the growth time (see Figure S5 in the Supplementary Information). It has been suggested that vdW growth on mica substrate for non-layered $Pb_{1-x}Sn_xSe$ and layered $VSe_2$ with strong interlayer interactions *via* a CVD method is of crucial importance to the thickness control, owing to stronger interaction between crystals and substrate. [19, 46] In this work, we have used various substrates including sapphire, glass and mica for growth of $PtTe_2$ single crystals. However, we find that the thicknesses of as-synthesized $PtTe_2$ single crystals are similar to those on $SiO_2$/Si substrate, indicating that the interlayer interactions of $PtTe_2$ are much stronger than that of the other materials.

Consistent with structure of $CdI_2$-type, synthesized sample shows Raman modes $E_g$ and $A_{1g}$ (see Figure 2c and d). [47] The intensity mappings show a high uniform contrast except for edges, indicative of high quality crystal. To investigate polarization-dependent properties, we measured the Raman spectra of $PtTe_2$ under un- and parallel- as well as cross-polarized configurations in Figure 2e. There are two Raman-active modes positioned at ~110.7 cm$^{-1}$ and ~157.2 cm$^{-1}$, corresponding to $E_g$ and $A_{1g}$ modes. [13] As displayed in Figure 2e and Figure S6, the $E_g$ ($A_{1g}$) mode vanishes with polarization angle α varying from 0 to 180° under parallel (cross) polarization configuration, while the $A_{1g}$ ($E_g$) mode remains unchanged. These distinctive features indicate polarization-dependent anisotropic behavior of as-grown $PtTe_2$. [48, 49] In addition to the polarization dependence of modes, temperature variation can shift the peaks of $A_{1g}$ and $E_g$. The temperature-dependent softening modes of $PtTe_2$ (see Figure S7 in the Supplementary Information) is due to anharmonic contributions induced by phonon-phonon interactions. [50-53]

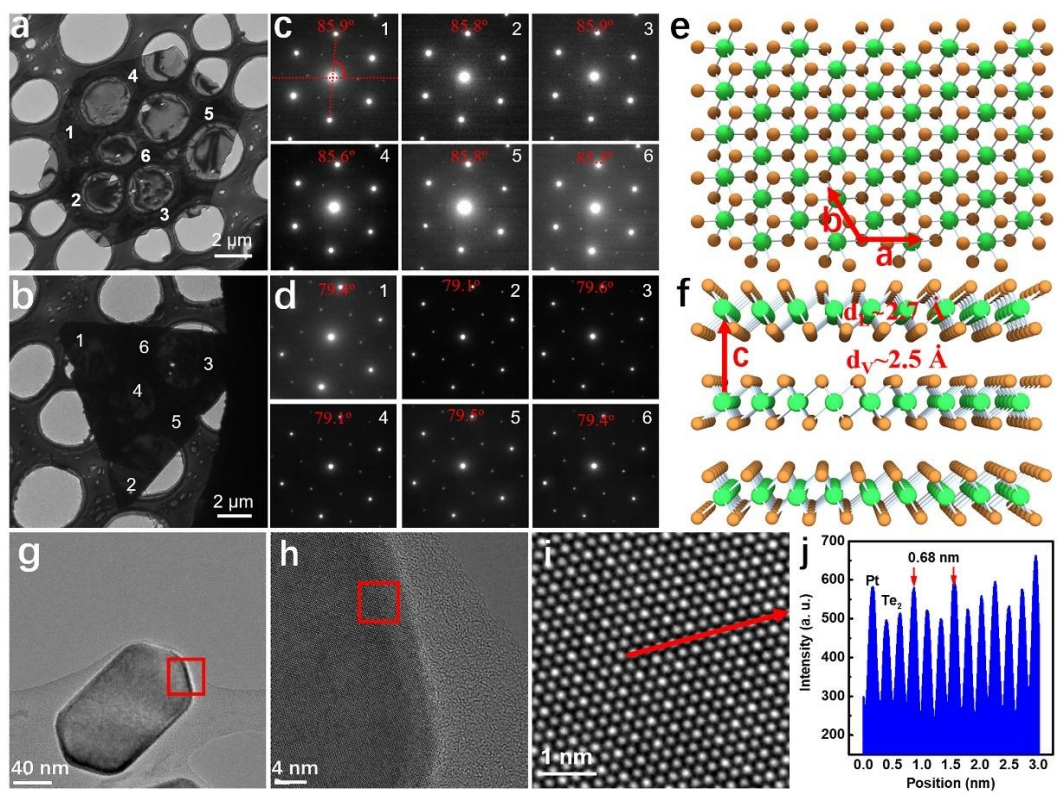

**Figure 3.** TEM characterizations of the as-grown $PtTe_2$ single crystals. (a-d) The low-magnification TEM images of hexagonal (a) and triangular (b) $PtTe_2$ flakes and their corresponding SAED patterns (c, d) taken from the points labeled with numbers 1-6, respectively. The angles of dashed lines linking the near vertical two dots with respect to the horizontal lines are marked in the corresponding patterns. (e, f) Top- and side-view structural models of $PtTe_2$ single crystal (Pt atoms, green; Te atoms, brown). (g-i) The lower magnification of $PtTe_2$ nanosheet (g) and corresponding zoomed-in higher magnification (h) and atomic-resolution TEM images (i) highlighted by red rectangles. (j) Intensity line profile extracted along the red solid line in (i).

Identification of the crystallinity and phase structure of as-grown PtTe$_2$ flakes is essential to reveal their unique physical properties. Figure 3a and b show low-magnification TEM images of typical hexagonal and triangular PtTe$_2$ crystals, respectively. To identify the crystallinity of the as-synthesized PtTe$_2$ flakes, a series of selected area electron diffraction (SAED) patterns at six different areas labeled with numbers 1-6 (see Figure 3c and d) were collected from the regular hexagonal- and triangle-shape PtTe$_2$. The angle deviation of less than 0.5° in all SAED patterns provides a solid proof of single crystallinity of SLS-synthesized PtTe$_2$ flakes. Compared to previous works that sulfurization or selenization of pre-deposited metals films on substrates yield TMDs polycrystalline films, [36, 54] our developed SLS approach based on a CVD-like setup can simply and rapidly synthesize high-quality TMDs single crystals. PtTe$_2$ as a typical CdI$_2$-type structure consists of one central Pt and six Te atoms located at octahedral points (see Figure 3e where green and brown balls stand for Pt and Te atoms, respectively). Bulk PtTe$_2$ can be viewed as the AA Bernal stacking of monolayer PtTe$_2$ (see Figure 3f). The most striking feature of PtTe$_2$ is that the vdW gap (2.5 Å) is smaller than the intrinsic layer thickness (2.7 Å), in sharp contrast to the other 2D materials. [33, 47] Such a smaller vdW gap indicates abnormally strong interlayer interactions, which yields relative thick PtTe$_2$ flakes in our SLS synthesis process.

To resolve the specific structure at atomic scale, we show atomic-resolution images in Figure 3i, by zooming in the area highlighted by red rectangle in Figure 3h. The alternative bright and dark atomic arrangement (Figure 3i) in the triangular lattice suggests that the as-synthesized PtTe$_2$ preserves the 1T stacking, where the bright and dark columns are Pt and Te atom chains, respectively. Based on the intensity line profile (see Figure 3j) along the red solid line in Figure 3i, the in-plane lattice constant (a) of our as-synthesized $PtTe_2$ is estimated to be 0.39 nm, in excellent agreement with the previous report. [55]

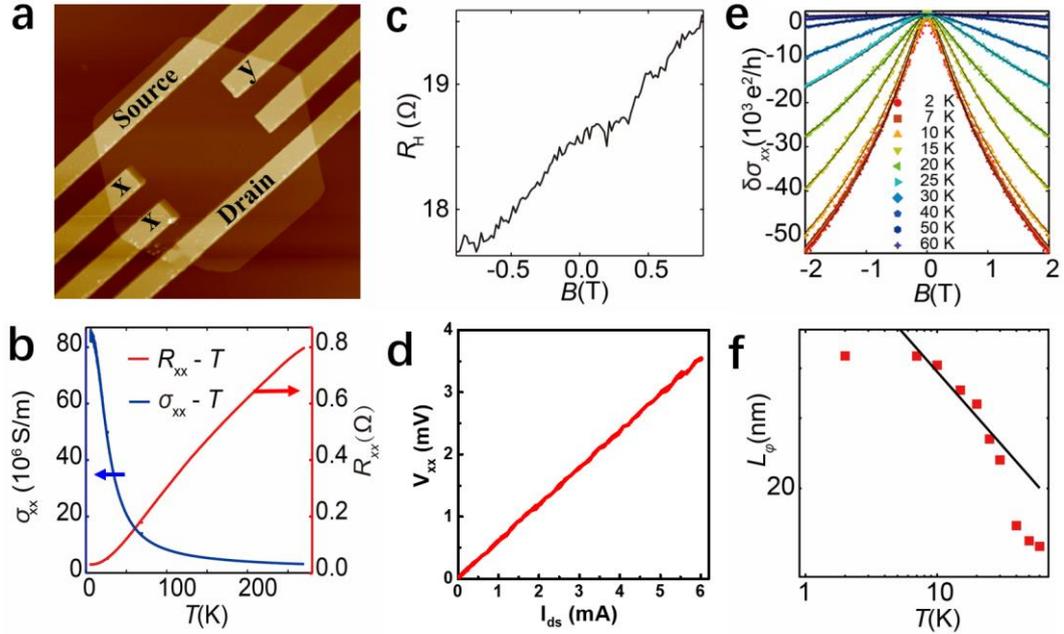

**Figure 4.** The electrical transport of SLS-grown $PtTe_2$ single crystal device. (a) The AFM topography image of a $PtTe_2$ Hall bar single crystal device. (b) $R_{xx} – T$ and the corresponding $\sigma_{xx}–T$ curves. $\sigma_{xx} = (L/WH)/R_{xx}$, where $L$, $W$ and $H$ are the length, width, and thickness of the channel, respectively. (c) The hall $R_{xy}$-B curve. (d) $V_{xx}$-$I_{ds}$ characteristic curve. (e) The temperature dependent magnetoconductivity $\delta\sigma_{xx}(B)–B$. The symbols represent experimental data and solid lines are theoretical results based on Eq. (1). (f) The extracted dephasing length $L_\varphi$ vs $T$ shows a power-law rule of $T^{-\gamma}$ (the black solid line) with the $\gamma=0.37 \pm 0.05$.

As one of type-II Dirac semimetals, $PtTe_2$ is supposed to own excellent electrical properties. We fabricated $PtTe_2$ Hall bar devices (see Figure 4a) and performed measurement of resistivity at various temperatures (see Figure 4b) to reveal the electronic state property of $PtTe_2$. Similar to graphene, resistivity decreasing with temperature shows a metallic state. The measured carrier concentration of $PtTe_2$ single crystal is around $6.3 \times 10^{22}$ cm$^{-3}$ at room temperature (see Figure 4c), in accordance with its metallic property. Surprisingly, the measurement of I-V characteristic shows that the corresponding 4-terminal conductivity $\sigma_{xx}$ of PtTe$_2$ Hall bar devices can be $3.3 \times 10^6$ S/m at room temperature (see Figure 4d), which is a record high among 2D metallic materials reported so far. [14] To highlight the electrical conductivity of PtTe$_2$, we listed electrical conductivity of other 2D metallic materials and conventional electrode-used metals in Table 1. The value of electrical conductivity of as-grown PtTe$_2$ is superior to the state-of-the-art 2D metallic material VSe$_2$ ($1.1 \times 10^6$ S/m). [14] Moreover, the electrical conductivity is even better than that of titanium commonly used as electrode. We repeated measurements of electrical conductivity on more than 30 devices and found that all devices show similar results. Given that PtTe$_2$ single crystals have such a large electrical conductivity and atomically smooth surface, it may be used as electrode for 2D semiconducting materials to improve contact properties. [19, 56] Note that all the measurements of electrical conductivity were done at room temperature. At low temperature, electrical conductivity of PtTe$_2$ is strongly dependent on specific scattering mechanisms.

To investigate how electron scattering mechanisms affect electrical conductivity of PtTe$_2$ at low temperatures, we carried out the measurement of 4-terminal electrical transport in a He$^4$ cryostat at various magnetic fields. Interestingly, we observe that the perpendicular magnetic field dependence of magnetoconductivity of PtTe$_2$ exhibits a WAL behavior (see Figure 4e), which persists up to 60 K and becomes much more prominent at lower temperature. The WAL effect can be manifested by the quantum interference induced change in electrical conductivity of diffusive regime. Besides, the presence of WAL effect is an indication of strong spin-orbit coupling in PtTe$_2$, which is consistent with the fact that heavy transition metal element Pt leads to a strong spin-orbit interaction in PtTe$_2$. In the limit of strong spin-orbit coupling, we fitted the experimental data of magneto-conductance (defined as $\delta\sigma_{xx}(H) = (L/WH)/R_{xx}(H) - (L/WH)/R_{xx}(0)$) using Hikami-Larkin-Nagaoka (HLN) model [57]

$$\Delta\sigma(B) = \frac{\alpha e^2}{\pi h}\left\{\left[\ln\left(\frac{B_\varnothing}{B}\right) - \Psi\left(\frac{1}{2} + \frac{B_\varnothing}{B}\right)\right]\right\}, \quad (1)$$

where $e$ is the elementary charge, $h$ is the Planck constant, $\psi$ is the digamma function, and $B_\varphi$ refers to the characteristic field of phase coherence. As shown in Figure 4e, the HLN theory (solid lines) agrees well with experimental data (symbols). Similar to the previous reports, α is much larger than that in 2D systems due to bulk effect. [58, 59] Through the fittings, we extracted the corresponding phase decoherence length $l_\varphi = \sqrt{h/(8\pi e B_\varphi)}$, with results shown in Figure 4f. It is found that $l_\varphi$ displays a $T^{-\gamma}$ dependence with γ=0.37 ± 0.05 within temperature ranges from T = 7 K to 30 K. The extracted value of γ indicates that electron dephasing in PtTe$_2$ at low temperatures may be attributed to electron-electron interactions (γ=0.5), rather than the electron-phonon interactions (γ=1).

**Table 1.** Comparison of the electrical conductivity for SLS-grown PtTe$_2$ single crystal, several 2D metallic materials and conventional metals.

| Different Materials | Conductivity (Sm$^{-1}$) | References |
|---|---|---|
| 1T-MoS$_2$ | 1.0-10 ×10$^3$ | [60] |
| rGO | 1.0 ×10$^4$ | [61] |
| Graphene | 0.2-2.0 ×10$^6$ | [62-64] |
| CVD-VS$_2$ | 3.0×10$^5$ | [19] |
| CVD-VSe$_2$ | 1.1 ×10$^6$ | [14] |
| Stainless Steel | 1.4 ×10$^6$ | [65] |
| Titanium | 2.4 ×10$^6$ | [66] |
| PtTe$_2$ | 3.3 ×10$^6$ | This work |
| Gold | 4.1×10$^7$ | [67] |

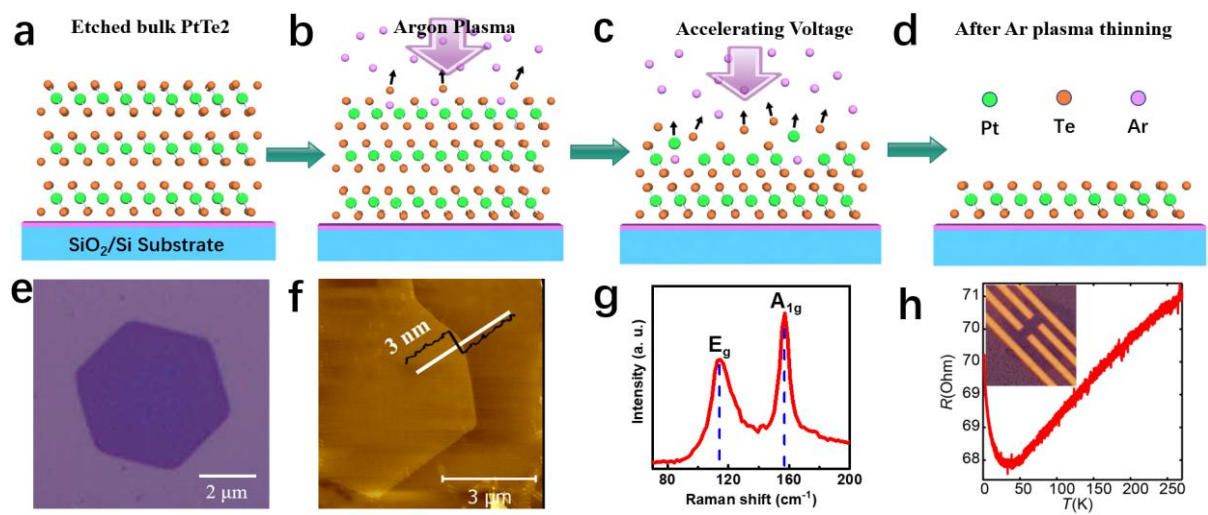

**Figure 5.** (a-d) Schematic showing the processes of layer-by-layer thinning of bulk PtTe$_2$ single crystals by accelerated argon plasma. (e, f) Optical microscopy and AFM topography images of PtTe$_2$ single crystal after argon plasma thinning, respectively. The inserted height profile extracted along the white solid line in (f) gives a thickness of ~3 nm. (g) Raman spectrum of PtTe$_2$ flake after argon plasma thinning. (h) Temperature-dependent resistivity for the thinned PtTe$_2$ device. The inset shows a typical Hall bar device fabricated with a hexagonal thinned PtTe$_2$.

Physical properties of 2D materials are strongly dependent on their thickness, especially for group-10 TMDs materials [34, 37] and InSe [68-71] with strong interlayer interactions. To obtain atomically-thin samples, mechical exfoliation or CVD method is usually used. However strong interalyer interactions and high melting-point transition metal pose challenges for obtaining atomically-thin $PtTe_2$. Here a simple and general approach was developed to obtain the atomically-thin $PtTe_2$ with controlled thickness. We thinned as-synthesized bulk $PtTe_2$ single crystals down to atomic thickness by utilizing an argon plasma ion beam etching (IBE) method, as displayed in Figure 5a-d. When argon plasma is switched on, the relative light Te atoms at the top surface would be removed (see Figure 5b), leaving many Te vacancies. With plasma bombardment of longer time, heavy Pt atoms at the top surface would be also removed, leading to removal of $PtTe_2$ uni-layer (see Figure 5c). Eventually, atomically-thin $PtTe_2$ single crystals can be obtained (see Figure 5d).

The obtained atomically-thin $PtTe_2$ crystals through argon plasma IBE method still retain high crystallinity. Figure 5e shows an optical microscopy image of a plasma-thinned equilateral hexagonal-shape sample with identical color contrast, which is primarily assigned to single crystalline $PtTe_2$ with uniform thickness. Surface roughness of less than 0.5 nm in the height profile indicates that the thinning approach of using argon plasma does not destroy its crystallinity and that few of $PtTe_2$ traces are left in the surface. We note that similar approach has been applied for bulk samples with weak interlayer interaction to obtain thin flakes, such as $MoS_2$ layer thinning with a local anodic oxidation method. [72-74] Different from prior works, we modified IBE instrument to speed up argon plasma and successfully thinned as-grown $PtTe_2$ with strong interlayer interaction down to atomically-thin crystals. To obtain high-quality $PtTe_2$ flakes through thinning approach, we carefully controlled the energy of argon plasma as small as possible by adjusting the accelerating voltage.

To further examine the crystallinity of post-thinning PtTe$_2$, we performed Raman spectroscopy and Raman mapping on the samples. Raman spectrum in Figure 5g shows two prominent E$_g$ and A$_{1g}$ modes centered at ~116.7 cm$^{-1}$ and ~157.0 cm$^{-1}$, consistent with that of pre-thinned PtTe$_2$ crystals. As expected, we find that the E$_g$ mode has blue-shift of 6.7 cm$^{-1}$ compared to as-synthesized bulk ones, due to strong interlayer interactions. [75, 76] The full width at half maximum (FWHM) of A$_{1g}$ peak is not sensitive to the variation in thickness of PtTe$_2$, therefore it could be used as a criteria to evaluate the quality of plasma-thinned PtTe$_2$ single crystals. [77] The FWHM of A$_{1g}$ mode is 6.9 cm$^{-1}$ and is close to as-synthesized bulk one (6.7 cm$^{-1}$), thus providing a further proof of high-quality PtTe$_2$ crystal after etching. To examine the change in electrical conductivity, we fabricated a Hall bar device (see the inset in Figure 5h) based on thinned PtTe$_2$ flake. Measurement of electric transport (see Figure 5h) demonstrates that thinned samples still retain a superior electrical conductivity (0.43×10$^6$ S/m), which suggests a huge potential of plasma etching technique in thinning layered materials with strong-interlayer interactions.

**3. Conclusion**

In summary, we realize the first low-temperature eutectic liquid-phase synthesis of high-quality PtTe$_2$ single crystals with thickness ranging from 2 to 200 nm in an ambient-pressure CVD system. The growth mechanism is based on solid-liquid-solid eutectic solidification, which is completely different from conventional high temperature gas-phase synthesis of TMDs. We find that the electrical conductivity of as-grown PtTe$_2$ is superior to all metallic TMDs ever reported. We also experimentally identify the weak antilocalization phenomenon in type-II Dirac semimetal for the first time. Furthermore, we develop a simple and general strategy of obtaining atomically-thin PtTe$_2$ samples with controllable thickness by lay-by-layer thinning as-grown samples, which can still retain high-crystallinity and exhibits excellent electrical conductivity. The approaches reported in this paper open up a new avenue for the controllable growth and fabrication of TMDs materials with high-melting-point transition metal and strong interlayer interactions, and paves the way for realizing novel optoelectronc and topological electronics based on 2D TMDs materials.

**4. Experimental Section**

**Material synthesis:** Tellurium (Te, Sigma-Aldrich, 4N purity) powders and Platinum (Pt, Sputtering target, 5N purity) films were used as the reactant precursors to synthesize PtTe$_2$ single crystals on SiO$_2$/Si substrates in the CVD-like furnace. The Ar/H$_2$ mixture was used as the carrier gas to carry Te vapor from the substrates to downstream. In a typical procedure of PtTe$_2$ growth, Pt films with various thicknesses (from 0.2 to 1.5 nm) were firstly deposited onto SiO$_2$/Si substrates as Pt reactant precursor using electron beam evaporation. Excess Te powder with different weight (from 50 to 200 mg) was uniformly dispersed on the top of the as-deposited Pt films. The synthesis processes were done in the CVD-like quartz tube with 1-inch diameter under ambient pressure. Upon pumping, the Ar/H$_2$ mixed gas flow at 500 sccm was introduced to purge air from the quartz tube for 20 mins, and then gas flow was maintained at 200 sccm during growth. The furnace was heated to 500 ºC in 20 mins, and then heated up to 700 ºC in 30 mins, followed by cooling down to 500 ºC in 30 mins controlled by a proportion integration differentiation (PID) controller. Finally, the furnace was naturally cooled down to room temperature with keeping Ar/$H_2$ gas flow and then switched off.

**Sample transfer and TEM characterization:** Polymethyl-methacrylate (PMMA) was spin-coated on the as-grown $PtTe_2$ flakes on $SiO_2$/Si substrates, and then PMMA/ $PtTe_2$ stack was transferred onto a TEM grid by etching $SiO_2$ in NaOH (2M) solution. Lastly, PMMA was removed by acetone and isopropanol. High-resolution Transmission Electron Microscopy (TEM, Titan 80-300) and Selected Area Electron Diffraction (SAED) were performed at the accelerating voltage of 200 kV.

**Material characterizations:** X-ray photoelectron spectroscopy was performed on a PHI 5000 Versaprobe system using AL Kα as X-ray source. The composition and elements distribution of as-synthesized flakes were determined by energy-dispersive X-ray spectroscopy attached to the scanning electron microscope (SEM, FEI Quanta 200 ESEM). The Atomic Force Microscopy measurements (Bruker multimode 8) were performed by using scanasyst mode. Raman spectroscopy/mappings were carried out under a 532.0 nm laser light and silicon-based CCD detector at room temperature using HORIBA JOBIN YVON HR800 Raman system. The spectra have been calibrated by 520.7 $cm^{-1}$ phonon mode from the silicon substrate. X-ray diffraction was used to measure the structural information of $PtTe_2$ crystals (Rigaku D/Max RAPID-II with Mo $K_α$ radiation, Japan).

**Device Fabrication and Electrical measurements:** A standard electron-beam lithography method was used to pattern (FEI F50 with Raith pattern generation system) electrodes. Ti/Au electrodes (5 nm/100 nm) were deposited *via* standard electron-beam evaporation. The $I_{ds}$-$V_{ds}$ curves of four-probe planar devices were measured by an Agilent B1500A semiconductor analyzer connected to a probe station at room temperature.

**Materials thinning:** A commercial Ion-Beam Etching instrument (Advanced IBE-150) with a Kaufmann ion source was utilized to thin as-grown $PtTe_2$ single crystals. The target holder was water-cooled and rotated at a constant velocity to ensure uniform etching. The base pressure was set to be higher than $8\times10^{-4}$ Pa before introducing the argon flow, and then maintained at $2\times10^{-2}$ Pa during etching process. Argon plasma with acceleration power of 90 eV was used with various current densities and irradiation time at room temperature.

**Supporting Information**
Supporting Information is available from the Wiley Online Library or from the author.
**Acknowledgements**
This work was supported in part by the National Key Basic Research Program of China (2015CB921600, 2016YFA0301204, 2013CBA01603), the National Natural Science Foundation of China (61625402, 11374142, 61574076, 11474147, 11474277 and 11434010), Fundamental Research Funds for the Central Universities (020414380093, 020414380084) and the Collaborative Innovation Center of Advanced Microstructures, the Key Research Program of the Chinese Academy of Sciences (Grant No. XDPB06-02, XDPB08-2). Dr. S. Hao would like to acknowledge the supports by China Postdoctoral Science Foundation (Grant No 2017M620203) and Postdoctoral Science Foundation of Jiangsu Province.
Received: ((will be filled in by the editorial staff))
Revised: ((will be filled in by the editorial staff))
Published online: ((will be filled in by the editorial staff))